\documentclass[a4paper,11pt]{article}
\usepackage{pos}
\usepackage{comment}
\usepackage{orcidlink}

\title{Minimal renormalizable $SO(10)$, spontaneous CP violation, and flavor implications}
\author*[\orcidlink{0000-0002-1361-4736}a]{Xiyuan Gao} 
\affiliation[a]{Institute for Theoretical Particle Physics, Karlsruhe Institute of Technology (KIT),\\
  Wolfgang-Gaede-Straße 1, D-76131 Karlsruhe, Germany}

\emailAdd{xiyuan.gao@kit.edu}
\abstract{In these proceedings, we summarize our recent findings on a minimal renormalizable $SO(10)$ grand unified theory. With the assumption of spontaneous $CP$ violation, the low-energy theory becomes a constrained two-Higgs-doublet model, whose mass spectrum has an upper bound of 545 GeV. High-luminosity collider experiments may find its flavor violating signals and reveal hidden mixing parameters, enabling new predictions for proton decay. There is a possibility that a hint of minimal $SO(10)$ can appear soon in near future experiments.}

\FullConference{9th Symposium on Prospects in the Physics of Discrete Symmetries (DISCRETE2024)\\
 2–6 Dec 2024\\
Ljubljana, Slovenia\\}

\begin{document}
\maketitle

\section{Introduction and motivation}
Grand unification theory (GUT)~\cite{Georgi:1974sy,Fritzsch:1974nn} is one of the most appealing frameworks for the new physics beyond Standard Model (SM). The SM fermions can fit neatly into the $\overline{5}$ and $10$ representation of $SU(5)$, or with one right-handed neutrino, into the $16$ representation of $SO(10)$. This does not appear to be a mere coincidence~\cite{Herms:2024krp}, and naturally explains the mysterious charge quantization~\cite{Foot:1992ui, Babu:1989tq}. 
GUT gives a profound prediction, proton decay. The proton lifetime is calculable, at least in certain minimal scenarios~\cite{Mohapatra:1979yj}.

Compared with $SU(5)$, $SO(10)$ is more attracting since it predicts massive neutrios. In addition, the intermediate scales enable successful gauge coupling unification, without requiring additional light states~\cite{Shafi:1979qb, Deshpande:1992au, Deshpande:1992em, Bertolini:2009qj}. If one insists renormalizability, a very minimal $SO(10)$ theory is based on large Higgs representations. The scalar sector contains~\cite{Bajc:2005zf, Bertolini:2012im, Preda:2025afo}
\begin{equation}
\label{large}
    45_H; \quad 126_H; \quad  \text{complexified}~10_H. 
\end{equation}
Clearly, the weakness is too many physical degrees of freedom, and perturbative expansion may fail~\cite{Milagre:2024wcg, Jarkovska:2023zwv}. Bearing this theoretical imperfection, perhaps the biggest shortcoming is that the renormalizable $SO(10)$ GUT itself lacks predictive power. Calculating all proton decay branching ratios is unfortunately not possible. Moreover, it does not require any new particles that could be tested in the upcoming colliders. There might be a particle desert all the way to an inaccessible high energy scale.

To find a predictive scenario while keeping renormalizability, we take a novel approach by imposing the condition that all couplings are real. CP symmetry, the combined transform of charge conjugation C and parity P, is enhanced and serves as a fundamental symmetry of nature~\cite{Landau:1957tp}. The CP-odd Kobayashi-Maskawa phase~\cite{Kobayashi:1973fv} derives only from the complex vacuum expectation values (VEV) for electroweak (EW) symmetry breaking, resulting in EW-scale spontaneous CP violation (SCPV).\footnote{High scale SCPV is not possible in the minimal theory. See Section~\ref{model} for an explanation.} Unlike the SM, the EW vacuua within SCPV must be degenerate, requiring additional fine-tuning and an extra light Higgs doublet~\cite{PhysRevD.27.1601, Nebot:2018nqn, Nierste:2019fbx}. The low-energy theory has no decoupling~\cite{Nierste:2019fbx} limit, very similar to what T.~Lee originally proposed in 1973~\cite{Lee:1973iz}. The phenomenology of the required new Higgs doublet would not only indicate sub-TeV new physics, but also shed light on additional flavour parameters. Related through $SO(10)$, those parameters are precisely the missing pieces needed to predict the proton decay branching ratios. 

In these proceedings, we summarize our recent work on minimal $SO(10)$ with SCPV~\cite{Gao:2024xte}. 
Section~\ref{model} introduces the model and describes the low energy theory. The predictions and conclusions are presented in Section~\ref{prediction} and~\ref{conclu}, respectively.

\section{SCPV in minimal $SO(10)$}
\label{model}
The fermionic sector for minimal $SO(10)$ is quite concise. All SM quarks, leptons, and three right-handed neutrinos are exactly contained in three spinor representations $16_F=(Q_L,u_R,d_R)+(\ell_L,\nu_R,e_R)$. The scalar sector, on the other hand, needs to be extended. As shown in Eq~\ref{large}, the simplest realistic choice for a renormalizable theory is $45_H, 126_H$ and complexified  $10_H$~\cite{Bajc:2005zf, Bertolini:2012im, Preda:2025afo}. Their transforming rules under CP can be defined as (with the $SO(10)$ indices implicit),
\begin{equation}
\label{cpdef}
    45_H\rightarrow 45_H, \quad 126_H\rightarrow \overline{126}_H, \quad 10_H\rightarrow 10^{*}_H. 
\end{equation}
This ensures that all couplings are real when the theory is exactly CP symmetric. The adjoint $45_H$ is a real. $126_H$ is complex, but its high-scale VEV $\langle (1,1,3,1) \rangle$ can always be chosen real by a phase redefinition~\cite{Bertolini:2012im}. Therefore, within the minimal $SO(10)$ scenario discussed here, the only possible VEV with a physical phase is the EW symmetry breaking one $\langle (1,2,2,0) \rangle$ from $10_H$ and/or $126_H$. CP is only spontaneously violated, together with EW symmetry breaking. The complete symmetry breaking chain is
\begin{equation}
    \begin{aligned}
        SO(10)\times CP ~&~\xrightarrow[M_{\text{GUT}}]{\langle (1,1,1,0) \rangle \in 45_H}SU(3)_C\times SU(2)_L\times SU(2)_R\times U(1)_{B-L}\times CP\\
        ~&~\xrightarrow[M_{R}]{\langle (1,1,3,1) \rangle \in 126_H}SU(3)_C\times SU(2)_L \times U(1)_Y \times CP
        \\
        ~&~\xrightarrow[M_{W}]{\langle (1,2,2,0) \rangle \in 126_H, 10_H} SU(3)_c\times U(1)_{\text{EM}}.
    \end{aligned}
\end{equation}
The $SU(3)_C\times SU(2)_L\times SU(2)_R\times U(1)_{B-L}$ quantum numbers are shown in parentheses. The intermediate symmetry can also be $SU(4)_C\times SU(2)_L\times U(1)_R\times CP$, while the other two-step Pati-Salam type breaking patterns are not possible here~\cite{Bertolini:2009qj, Ferrari:2018rey}. The gauge coupling unification will work perfectly~\cite{Deshpande:1992au,Deshpande:1992em, Bertolini:2009qj} when the intermediate LR (or QL) symmetry breaking scale $M_I$ is about $10^{9}$ GeV (or $10^{11}$ GeV). However, we do not specify any value, because the choice of $M_I$ can be relaxed if some of the physical states in $126_H$ and/or $45_H$ are fine-tuned light~\cite{Preda:2025afo}.

%If CP symmetry is spontaneously broken at EW scale, the EW vacuua must then be degenerate. This leads to domain wall solutions, a disaster for cosmology~\cite{Zeldovich:1974uw}. Fortunately, if the CP symmetry is not fully exact, a tiny CP-odd perturbation (often called biased term) can solve the domain wall problem. The domain wall network is unstable and collapses quickly after its formation~\cite{Zeldovich:1974uw, Vilenkin:1981zs, Gelmini:1988sf}. We assume the biased term derives the beyond grand unification physics, such as quantum gravity. It breaks CP explicity but has only negligible effects at low energy, except for destabilizing domain walls. So CP is phsically still a good symmetry even with the tiny perturbation. The non-zero KM phase, still dominantly originates from spontaneous CP violation, the complex EW VEVs. 

The SM EW symmetry breaking requires one VEV, one light Higgs doublet, and fine-tuning once. Correspondingly, EW scale SCPV necessities two degenerate VEVs\footnote{The degenerate vacuua leads to domain wall solutions, which is a disaster for cosmology~\cite{Zeldovich:1974uw}. It can be solved by a biased term~\cite{Zeldovich:1974uw, Vilenkin:1981zs, Gelmini:1988sf}.}, two light Higgs doublets, and double fine-tuning. This is an model-independent result, known for a long time~\cite{PhysRevD.27.1601}. But only until recently, it has been realized that the mass spectrum for these Higgs doublets is strictly bounded from above, regardless the size of the CP violating effect~\cite{Nebot:2018nqn, Nierste:2019fbx,Miro:2024zka}. This leads to a non-decoupling theory~\cite{Nierste:2019fbx}, with a perturbative unitarity bound in analog to the Lee-Quigg-Thacker one of SM~\cite{Lee:1977yc}
\begin{equation}
\label{illuUpperbound}
    m_h^2+m_H^2+m_A^2 ~\lesssim~ M_{\text{LQT}}^2 ~=~ (700\sim 800~\text{GeV})^2. 
\end{equation}
Here, $h$ as the discovered 125 GeV SM Higgs boson, and $H, A$ are the physical neutral states of two-Higgs-doublet models (2HDM)~\cite{Branco:2011iw}. Complete next-to leading order analysis tells that $m_H$ and $m_A$ are individually bounded by 485 GeV and 545 GeV, and the mass of the physical charged state should be smaller than 435 GeV~\cite{Nierste:2019fbx}.

At tree-level, $H$ and $A$ directly couple to SM charged Ferimons with
\begin{equation}
\label{YukawaPhys}
    \begin{aligned}
        -\mathcal{L}_{\Phi\overline{F}F}~\supset&~Y_E^{\ell\ell'} (H+i A)\overline{\ell_L} \ell'_R 
          +Y_D^{q q'} (H+i A)\overline{d^{q}_{L}} d^{q'}_{R}  
          +Y_U^{q q'} (H+i A)\overline{u_L^{q}} u_R^{q'} +\text{h.c.}
    \end{aligned}
\end{equation}
The flavor changing neutral currents (FCNC) arise here. Since $m_H, m_A$ are at the EW scale, the Yukawa coupling $Y_E, Y_D, Y_U$ can not be too large. Fortunately, we find the suppressed Yukawa couplings do not conflict with minimal $SO(10)$. On the other hand, the flavor structures are constrained at tree level. Noticing the $SO(10)$ Yukawa sector is
\begin{equation}
\label{YukawaSO10}
    \begin{aligned}
        -\mathcal{L}_Y ~=~ Y_{10} 16_F 10_H 16_F  + \widetilde{Y}_{10} 16_F 10_H^* 16_F  + Y_{126} 16_F \overline{126}_H 16_F  + \text{h.c.}
    \end{aligned}
\end{equation}
Here, $Y_{10}, \widetilde{Y}_{10}$ and $Y_{126}$ are $3\times3$ symmetric matrices at GUT scale~\cite{Mohapatra:1979nn}. All charged leptons and quarks directly get masses from Eq~\ref{YukawaSO10}, so their mass matrices are linear combinations of $Y_{10}, \widetilde{Y}_{10}$ and $Y_{126}$. Consequently, they are all symmetric and can be diagonalized with
\begin{equation}
\label{FermionMixing}
\begin{aligned}
    M_D~=~ D^{*}m_D D^{\dag}&, \quad
    M_U~=~ U^{*}m_U U^{\dag}, \quad 
    M_E~=~ E^{*}m_E E^{\dag};  \\
    V_{\text{CKM}}&~=~U^{\dag}D,  \qquad V_{E} ~=~ E^{\dag}D .    
\end{aligned}
\end{equation}
The only unknown flavor mixing matrix is $V_E$, the misalignment between the charged lepton and down-type quark sector. Furthermore, since $Y_E, Y_D, Y_U$ also originate from Eq~\ref{YukawaSO10}, they must be linear combinations of $Y_{10}, \widetilde{Y}_{10}$ and $Y_{126}$, too. As a result, $Y_E, Y_D, Y_U$ can reduce to linear combinations of $M_D, M_U, M_E$ and the flavor structure is significantly simplified. Ignoring the small $12$ and $13$ mixing angles in $V_{\text{CKM}}$, and the small masses of the first two generations, we arrive at
\begin{equation}
\label{NonDiag}
      Y_E^{\ell\ell'}~\propto~V_E^{\ell b}V_E^{\ell'b},\quad \ell\neq\ell';\qquad 
      Y_D^{q q'}~\propto~V_E^{\tau q}V_E^{\tau q'},\quad q\neq q'.\\
\end{equation}
Under these approximations, the Yukawa couplings for low-energy 2HDM follow next-to minimal flavor violation (NMFV)~\cite{Agashe:2005hk}.

\section{Prediction}
\label{prediction}
\begin{figure}[t]
  \centering
  \includegraphics[width=0.50\textwidth]{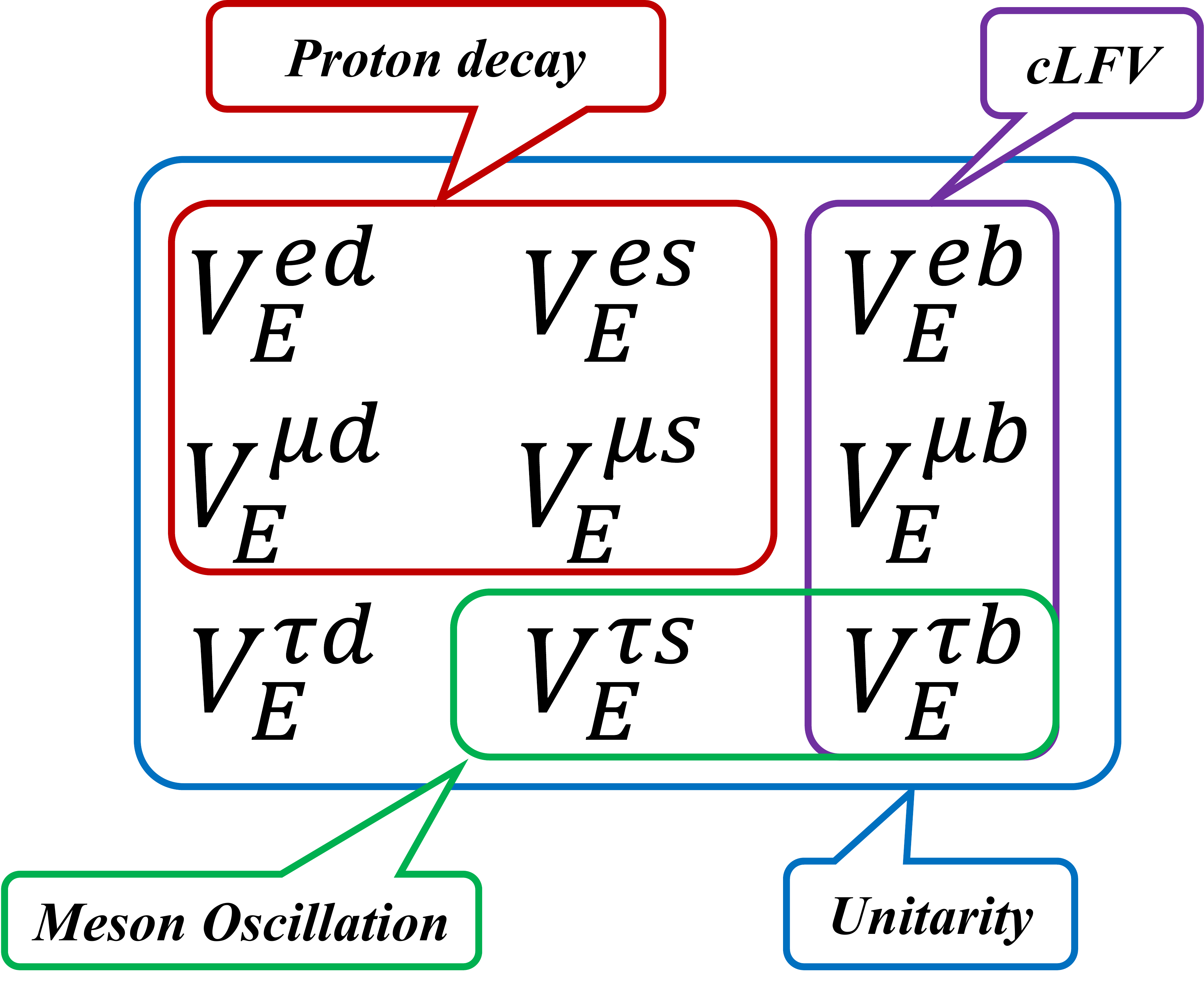}
    \caption{Proton decay, charged lepton flavor violation (cLFV), and neutral meson oscillation depend on different elements of the unitary matrix $V_E$. The theory yields more observables than free parameters.}
   \label{VE}
\end{figure}
If $A$ and $H$ are directly produced at colliders, signals of heavy resonances decaying into $e\mu, e\tau$ and $\mu\tau$ final states will be detected.  Additionally, the non-diagonal parts of $Y_D$ contribute to neutral meson oscillations. This means the SM predicted values of $K$ and $B_d$ meson oscillation frequencies should not perfectly match the measured ones. Moreover,  $V_E$ the misalignment mixing matrix between charged leptons and quarks, also joins the expressions of the proton decay branching ratios~\cite{FileviezPerez:2004hn, Nath:2006ut}. The dependence of these physical processes on different elements of $V_E$ is illustrated in Figure~\ref{VE}. Altogether, there are more physical observables than free parameters, and our concrete tree level prediction is
\begin{equation}
\label{expConnections}
\begin{aligned}
     &\frac{2\Gamma(p\xrightarrow{}\pi^0 \ell^+)}{\Gamma(p\xrightarrow{}\pi^+\overline{\nu})} -\frac{\Gamma(p\xrightarrow{}K^0 \ell^+)}{\xi_K\Gamma(p\xrightarrow{}K^+\overline{\nu})}\\
     ~=~&\left( \frac{3|(\Delta M_K)_{\text{NP}}|}{(\Delta M_{B_d})_{\text{NP}}|\xi_B}  - \frac{N_{e\mu}}{N_{\tau\mu}}-\frac{N_{e\mu}}{N_{e\tau}}+1\right)\left( \frac{N_{e\mu}}{N_{\tau\mu}}+\frac{N_{e\mu}}{N_{e\tau}}+1\right)^{-1}.
\end{aligned}
\end{equation}
Eq~\ref{expConnections} only contains experimental or lattice observables. $\Gamma(p\xrightarrow{}\pi^0 \ell^+)$ and $\Gamma(p\xrightarrow{}K^0 \ell^+)$ represents the total decay rate for all leptonic final states ($e^+$ and $\mu^+$). $N_{\ell\ell'}$ is the total number of the excess events of heavy resonant decaying to charged leptons, normalized by the detection efficiency. $(\Delta M_K)_{\text{NP}}$ and $(\Delta M_{B_d})_{\text{NP}}$ is difference between the SM predicted and experimentally measured values of $K$ and $B_d$ meson oscillation frequency. 
$\xi_B\approx0.97$ and $\xi_K\approx6.4$ are ratios between hadronic elements. They are well-calculated and one can find the explicit expressions in~\cite{Gao:2024xte}. Verifying Eq~\ref{expConnections} in future experiments would directly hint minimal $SO(10)$. Otherwise, the minimal theory does not hold or the one-loop corrections are overlooked.

\section{Conclusion and discussion}
\label{conclu}
In these proceedings, we have discussed the minimal realistic $SO(10)$ under the assumption of spontaneous CP violation. The low-energy theory is 2HDM and very similar to T. Lee's original proposal~\cite{Lee:1973iz}. One may worry the dangers tree level FCNC of the theory, but strictly speaking, the theory can not be trivially excluded. The absolute strengths of the Yukawa couplings are are not faithfully predicted in general, neither within the minimal $SO(10)$. What to be predicted is the flavor structure. Future measurement on the induced FCNC processes may reveal a hidden flavor mixing matrix $V_E$. That's exactly the missing piece to predict the proton decay branching ratios. A tree-level correlation is explicitly shown in Eq~\ref{expConnections}. Hopefully in near future, the high-luminosity LHC and Hyper-Kamiokande experiments can provide a hint for $SO(10)$.

\section*{Acknowledgments}
The author thanks Goran Senjanović, Ulrich Nierste, and Robert Ziegler for useful discussions on this project. This research was partly supported the BMBF grant 05H21VKKBA, \textit{Theoretische Studien für Belle II und LHCb.} X.G. also acknowledges the support by the Doctoral School ``Karlsruhe School of Elementary and Astroparticle Physics: Science and Technology''.

\bibliographystyle{JHEP}
\bibliography{SO10_CPV_ref.bib}

\end{document}